\documentclass[manuscript]{acmart}

\AtBeginDocument{%
  }

\setcopyright{acmlicensed}
\copyrightyear{2026}

\usepackage{graphicx}
\usepackage{subcaption}
\usepackage{url}
\usepackage{breakurl}

\begin{document}

\acmConference[XAIxArts 2026]{Explainable AI for the Arts Workshop 2026}{July 13, 2026}{London, UK}

\acmDOI{}
\acmISBN{}

\title{Still searching for an \textit{(un)stable equilibrium}: visualising the process of training generative neural networks without data}


\author{Terence Broad}
\email{t.broad@arts.ac.uk}
\orcid{0000-0001-9987-6536}
\affiliation{%
  \institution{Creative Computing Institute, University of the Arts London}
  \country{United Kingdom}
}

\renewcommand{\shortauthors}{Broad}

\begin{abstract}

\textit{(un)stable equilibrium} is an ongoing series of works that is based on a practice of training generative neural networks without data. This paper introduces the second series of \textit{(un)stable equilibrium} works, in which the process of training without data is visualised in a series of video pieces. These works show a generative network attempting to converge to a fixed point that is undefined, caught in an endless, unresolvable search for equilibrium. The video pieces described in this paper guide the viewer toward an understanding of AI through aesthetic experience rather than technical exposition, and strive to give a conceptual understanding of what AI could be, rather than a restatement of what it currently is. This project sits within a broader set of artistic practices that serve as an alternative and critical modes of explainability for AI.

\end{abstract}


\maketitle

\section{Introduction}

\textit{(un)stable equilibrium} is a series of artworks that was first created in 2019 \citep{broad2019searching}. From the outset it was conceived as an ongoing practice comprising multiple series of works. This paper presents describes the technical implementation of the second such series of \textit{(un)stable equilibrium} works.

In developing this practice, I set myself a number of constraints that govern and inform the creative process \cite{candy2007constraints} by which the works in this series are produced: All networks\footnote{The term network is deliberately used here instead of model, as a generative model, is a statistical \textit{model} of the dataset it has been trained on. As there is not data used for training, the resulting network is not a model of anything in the formal sense.} used in training must be randomly initialised, and no external data may be used in the training of any network, nor can external data be used indirectly in the training process or be given as inputs to the network. No pretrained model that has been trained on an external dataset is permitted to be used in the training process. Similarly, no external mathematical model or representation of aesthetic value may be used to guide the training process. Instead, any data used within the loss function must emerge entirely from the dynamics of the network training, or dynamics between the various networks in the training ensemble.

While I am not the only practitioner to have used networks not trained on data, alternative approaches tend to use randomly initialised networks without any training \citep{chelma2023genesis} or make use of some kind of external representation derived from data as inputs to generative networks where the outputs are still undefined \cite{simon2019dimensions,park2020generating}. Existing approaches also make use of fine-tuning a network whose representations were learned from the dataset on which the original network was trained \citep{broad2020amplifying} or from an external dataset \cite{broad2019transforming,gal2022stylegan}.

By rejecting any form of external data, the approach to creating works in this series is more akin to practices in traditional generative art, where dynamic systems are built and the role of the artist is to design or influence this process to some degree, based on intuition and exploration. The only difference being is that the tools being used are gpu-optimised linear algebra libraries, differentiable objective functions and gradient-based optimisation. In this approach the tools for machine learning are not creating statistical models of data per se, but rather co-opting them to produce art using statistical means, which Bense refers to as creating new `aesthetic structures' \citep{bense1965projekte}. This fits into a broader body of artistic practices that craft small models \cite{abuzuraiq2024seizing} or use AI networks and training tools as artistic materials \cite{broad2024using}, where artistic interventions into the normal functioning of AI lead to the creation of artworks that help to serve as a form of unconventional and critical explanation of AI \cite{broadforthcomingexplaining}.

\section{\textit{(un)stable equilibrium}: Series 1}

The original series of works in the \textit{(un)stable equilibrium} was first described in \cite{broad2019searching} and later expanded on in \cite{broad2025expanding}. The arrangement used for training is a modified version of the generative adversarial networks (GAN) framework \citep{goodfellow2014generative}. Where there are two Generator networks $G_{1}$ \& $G_{2}$. Under this formulation, the Discriminator $D$ attempts to correctly distinguish between the outputs of each Generator, while each Generator attempts to imitate the other:




\begin{equation}
\min_{G_{1}}\max_{G_2}\max_{D}\mathbb{E}_{z\sim p_{\text{z}}(z)}[\log{D(G_{2}(z))}] + [1 - \log{D(G_{1}(z))}]
    \label{eq:double-gen-gan-loss}
\end{equation}

In addition to this, another loss term is also used to train the generators, in which each is optimised to produce greater colour variation in its output batch of generations than the other network. This colour variation loss term is defined as:

\begin{equation}
    \label{eq:variance-gan}
    Vdiff = Var(B_{g_{1}}^{c}) - Var(B_{g_{2}}^{c})
    \end{equation}

\begin{figure}[!htbp]
    \centering
    \subfloat[]{\label{subfig:ue_1_1}\includegraphics[width=0.31\textwidth]{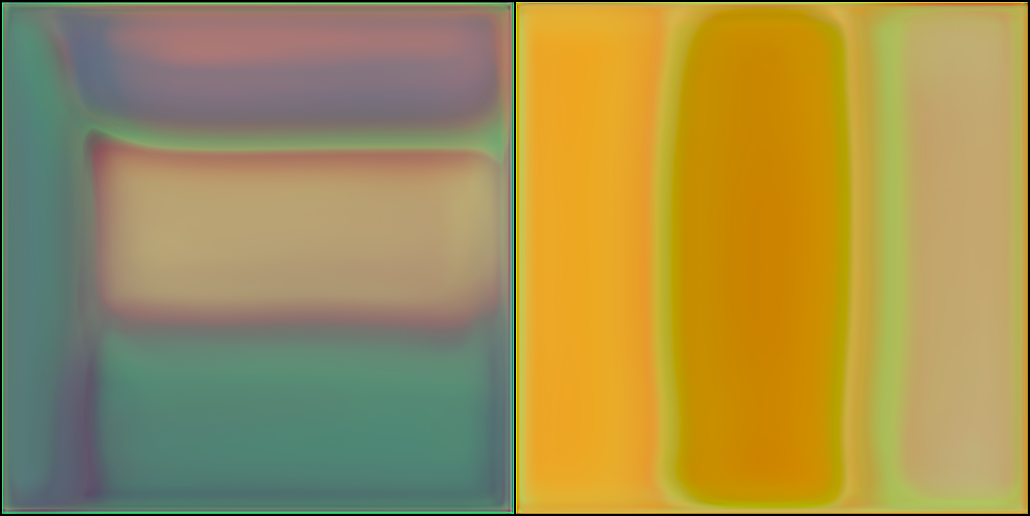}}
    \hfill
    \subfloat[]{\includegraphics[width=0.31\textwidth]{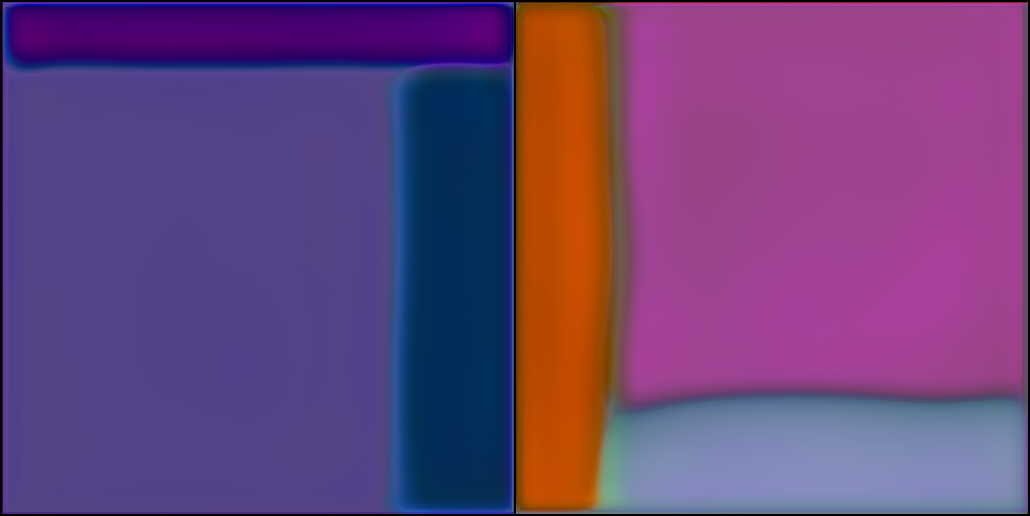}}
    \hfill
    \subfloat[]{\includegraphics[width=0.31\textwidth]{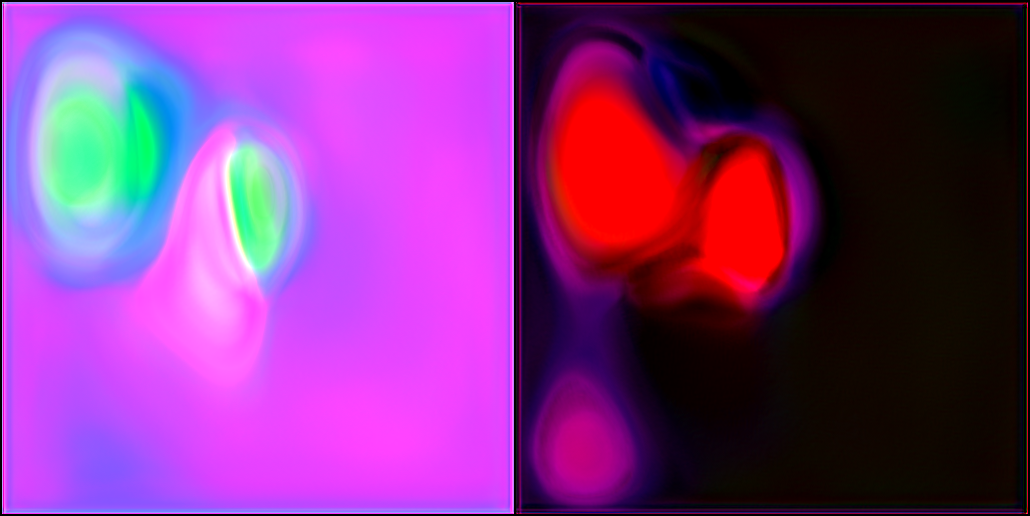}}
    \hfill
    \subfloat[]{\includegraphics[width=0.31\textwidth]{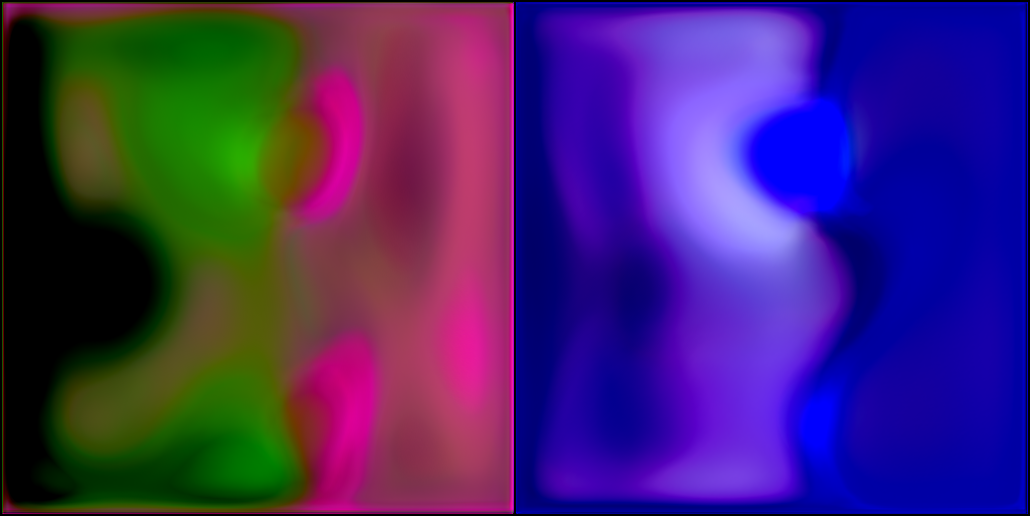}}
    \hfill
    \subfloat[]{\includegraphics[width=0.31\textwidth]{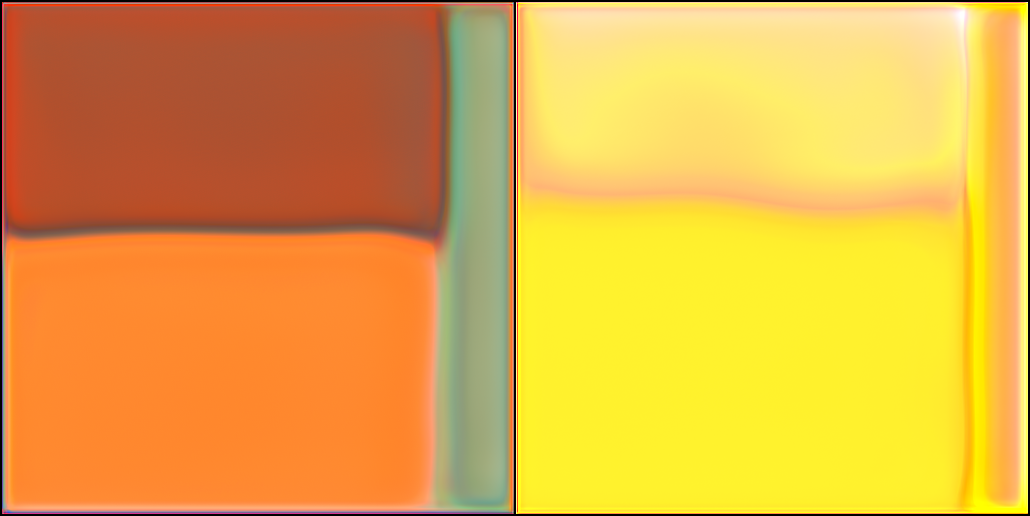}}
    \hfill
    \subfloat[]{\includegraphics[width=0.31\textwidth]{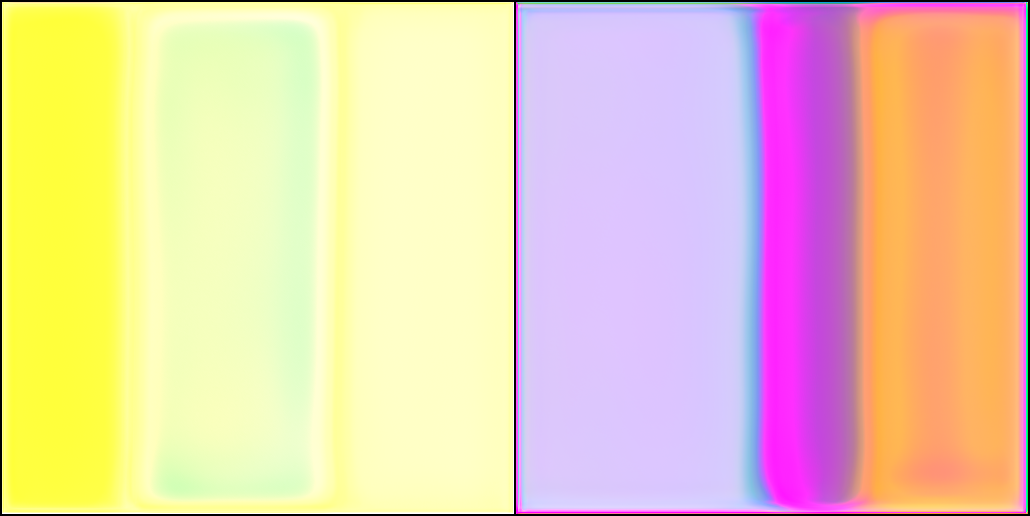}}
    \caption{Stills from the works in the original \textit{(un)stable equilibrium} series. (a) \textit{(un)stable equilibrium 1:1}, (b) \textit{(un)stable equilibrium 1:2}, (c) \textit{(un)stable equilibrium 1:3}, (d) \textit{(un)stable equilibrium 1:4}, (e) \textit{(un)stable equilibrium 1:5}, (f) \textit{(un)stable equilibrium 1:6}.}
    \label{fig:original-experiments}
  \end{figure}

The result of this training procedure was the artwork \textit{(un)stable equilibrium 1:1} (Fig. \ref{subfig:ue_1_1}). Five further experiments were conducted in this first series (Fig. \ref{fig:original-experiments}b-f), all trained with variations on the original adversarial loss given in Eq. \ref{eq:double-gen-gan-loss}.

\section{Fine-tuning with inverse adversarial loss}

The basis of the experiment used to make the second series of works in \textit{(un)stable equilibrium} (\S \ref{sec:4}), is based on prior work originally for fine-tuning pre-trained models without data. In \cite{broad2020amplifying} I describe an experiment in which I performed fine-tuning towards `unlikelihood' on a pretrained StyleGAN trained on the Flickr Faces High Quality (FFHQ) dataset \citep{karras2019style}. This was fine-tuned in a divergent fashion \citep{broad2021active}, away from the likelihood of producing realistic faces and towards a prediction of `unlikelihood' given by the frozen weights of the Discriminator $D_{f}$. The loss function for this fine-tuning procedure is:

\begin{equation}
  \max_{G}\mathbb{E}_{z\sim p_{\text{z}}(z)}[1 - \log{D_{f}(G(z))}]
  \label{eq:inverted-adv-loss}
  \end{equation}

\begin{figure}[!htbp]
    \subfloat[]{\label{subfig:og-samples}\includegraphics[width=.48\textwidth]{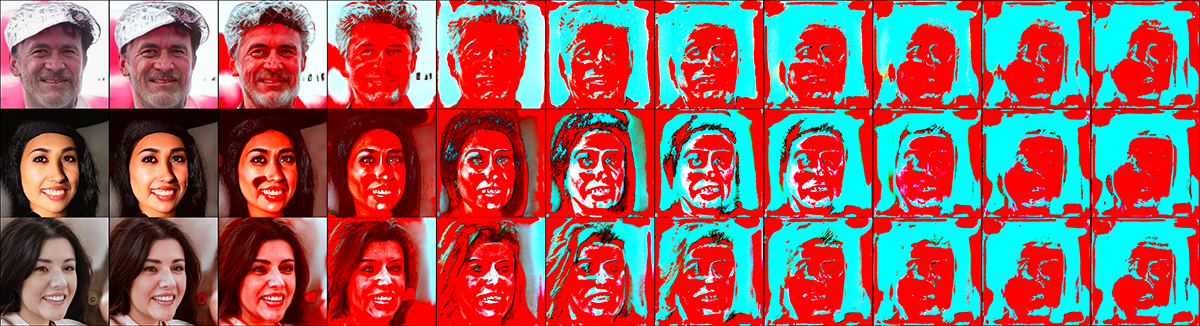}}
    \hfill
    \subfloat[]{\label{subfig:og-loss}\includegraphics[width=.48\textwidth]{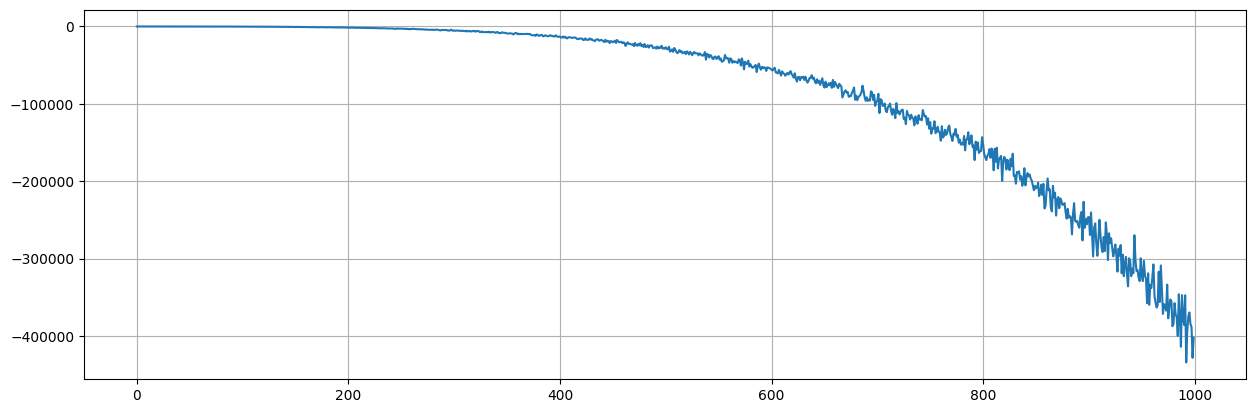}}
    \hfill
    \caption{(Generated samples (a) and loss plot (b) from the fine-tuning procedure given by Equation \ref{eq:inverted-adv-loss}, samples taken increments of 100 iterations, between training steps 0-1000.}
    \label{fig:og-loss}
 \end{figure}

This fine-tuning process causes the gradient to explode into large negative numbers while the visual outputs of the network collapse into a single representation (Fig. \ref{fig:og-loss}). When I revisited these experiments in preparing my PhD thesis \cite{broad2025expanding}, I re-ran them with a simple modification intended to mitigate this exponential explosion of gradients, which was to take the natural logarithm (aka the log) of the loss given in equation \ref{eq:inverted-adv-loss}:

\begin{equation}
  \max_{G}\mathbb{E}_{z\sim p_{\text{z}}(z)}[\log(1 - \log{D_{f}(G(z))})]
  \label{eq:ln-inverted-adv-loss}
  \end{equation}

 \begin{figure}[!htbp]
    \subfloat[]{\label{subfig:log-samples}\includegraphics[width=.48\textwidth]{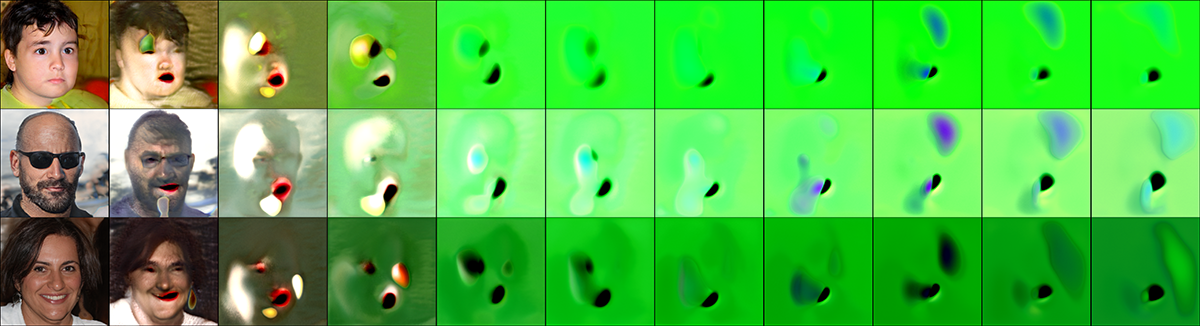}}
    \hfill
    \subfloat[]{\label{subfig:log-loss}\includegraphics[width=.48\textwidth]{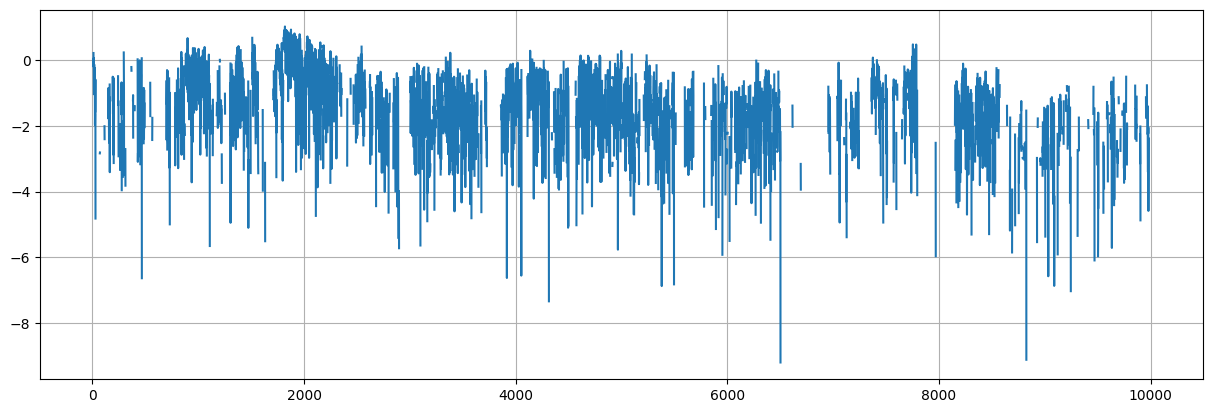}}
    \hfill
    \caption{Generated samples (a) and loss plot (b) from the fine-tuning procedure given by Equation \ref{eq:ln-inverted-adv-loss}, samples taken increments of 1000 iterations, between training steps 0-10000.}
    \label{fig:log-loss}
 \end{figure}

The unintended consequence of this intervention was to produce an objective that optimises towards something that is mathematically undefined. As this loss tends negative, and the log of a negative number is undefined, the gradient descent training process therefore optimises toward a point in loss space that can never be reached (Fig. \ref{fig:log-loss}). Training therefore continues in perpetuity, never resolving. This training approach became the basis of the second series of \textit{(un)stable equilibrium}.

\section{\textit{(un)stable equilibrium}: Series 2}
\label{sec:4}

Though initially used for fine-tuning a pretrained model, the training process described in the previous section by Eq. \ref{eq:ln-inverted-adv-loss} fits all of the criteria that was set out in the original constraints for creating the \textit{(un)stable equilibrium} artworks. I therefore made use of this method with one small modification to create the second series of \textit{(un)stable equilibrium} works. Rather than fine-tune an pretrained model, this loss is used to train a network entirely from scratch. The objective remains the same: maximising the generator's output with respect to the prediction of `fake' by the frozen weights of the discriminator. In this case, however, both the generator and discriminator are randomly initialised networks.

 \begin{figure}[!htbp]
    \subfloat[]{\label{subfig:ue_2_1_training}\includegraphics[width=1\textwidth]{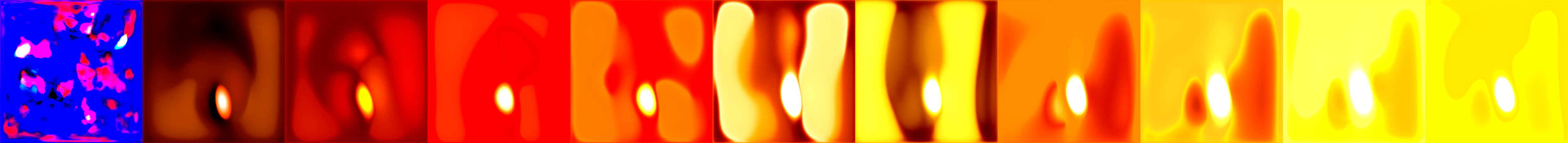}}
    \hfill
    \subfloat[]{\label{subfig:ue_2_2_training}\includegraphics[width=1\textwidth]{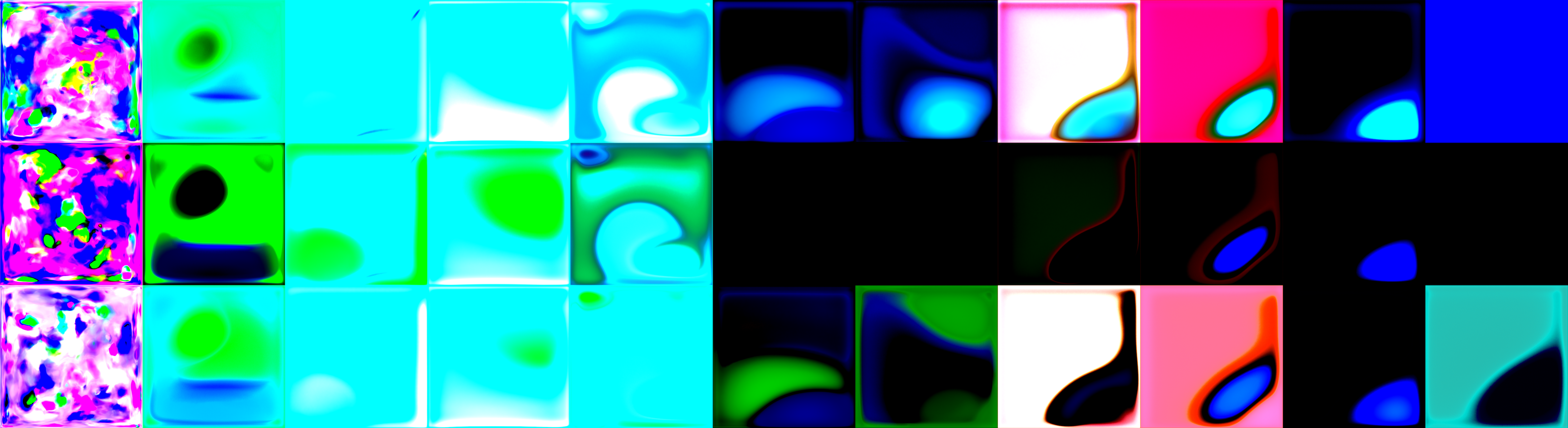}}
    \hfill
    \caption{The sequence of training for both the video works \textit{(un)stable equilibrium 2:1} (a) and \textit{(un)stable equilibrium 2:2} (b). Samples are taken at increments of 10000 iterations, between 0 and 100,000 iterations.}
    \label{fig:ue_2_training}
 \end{figure}

\begin{figure}[!htb]
    \centering
    \captionsetup{justification=centering}
    \includegraphics[width=0.33\textwidth]{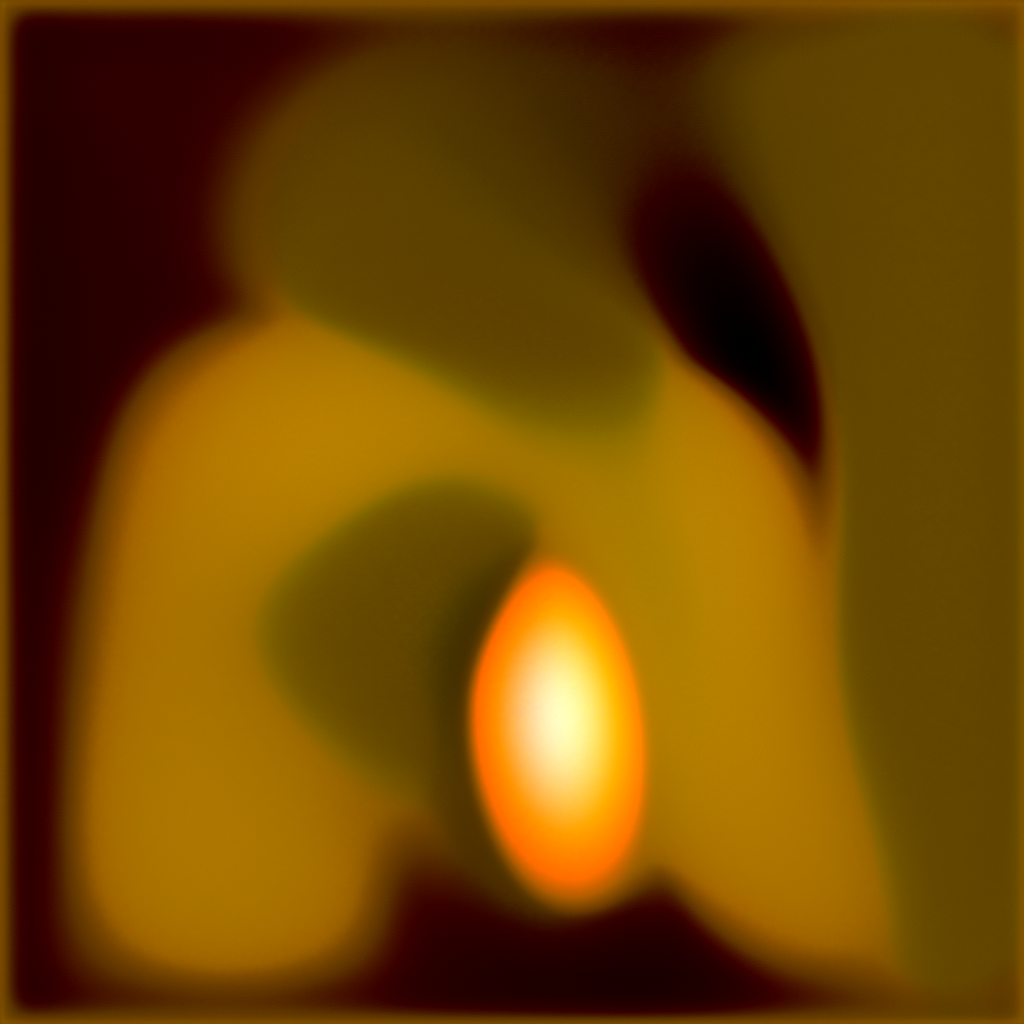}
    \caption{Still from \textit{(un)stable equilibrium 2:1}. Full video available at: \url{https://www.youtube.com/watch?v=YCGOCrLmMvQ}}
    \label{fig:ue_2_1_still}
\end{figure}

\begin{figure}[!htb]
    \centering
    \captionsetup{justification=centering}
    \includegraphics[width=1\textwidth]{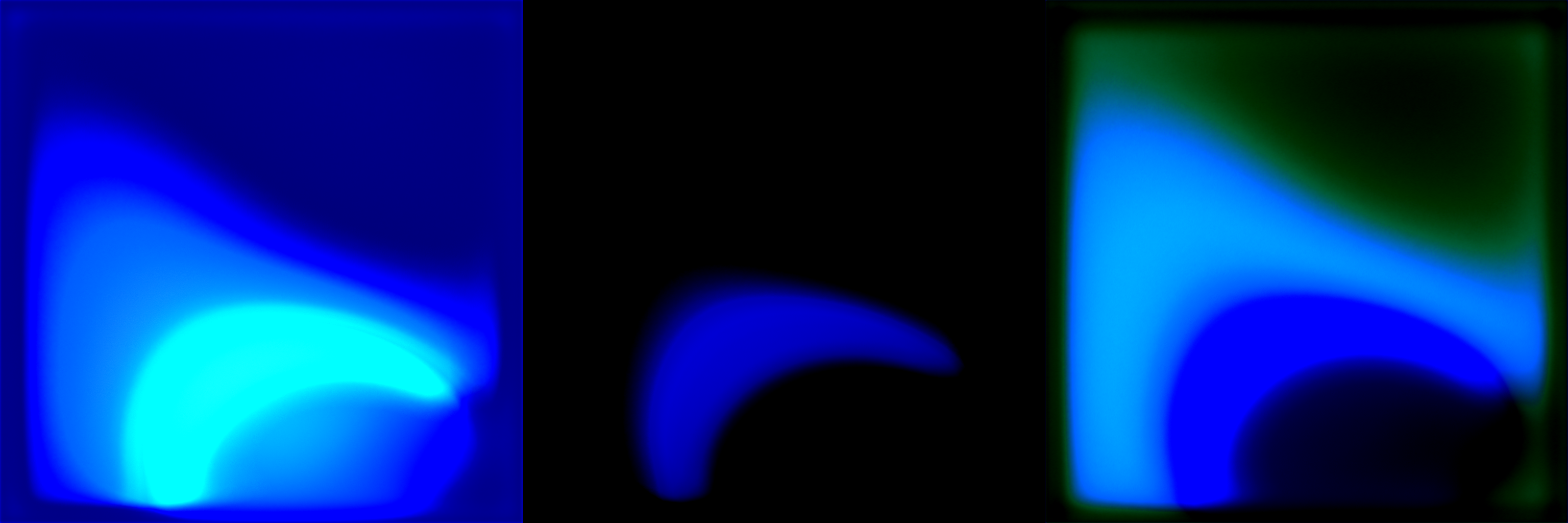}
    \caption{Still from \textit{(un)stable equilibrium 2:2}. Full video available at: \url{https://www.youtube.com/watch?v=1vjVn7diPSE}}
    \label{fig:ue_2_2_still}
\end{figure}

As with the previous \textit{(un)stable equilibrium} series, the body of work comprises a number of artworks, each resulting from a separate training experiment. Unlike the first series, in which video works presented a latent space interpolation computed after training was complete, these works document the process of training itself (as can be seen in Fig. \ref{fig:ue_2_training}). In \textit{(un)stable equilibrium 2:1} (Fig. \ref{fig:ue_2_1_still}), an image is generated at each training iteration from a fixed latent vector; once training is complete, these images are sequenced into a video that documents the unfolding training process. In the second experiment in the series, \textit{(un)stable equilibrium 2:2} (Fig. \ref{fig:ue_2_2_still}), the same procedure is repeated with images generated from three fixed latent vectors at each training step. These three image sequences can be presented together as a single video work, or as three separate synchronised video channels. Though only two works in this series have been created thus far, further experiments of greater multi-channel videos are planned, as well as other variations of losses and network architectures used as well as potentially creating an interactive version of this training procedure.


\section{Discussion}

\textit{(un)stable equilibrium} was originally named as such, because the artistic process behind the creation of the works was a long search for a configuration for training that would find an equilibrium in the space of potential system dynamics which is stable enough to prevent gradients collapsing or exploding, but unstable enough to produce unexpected results. The works in the second series, through animating the process of training towards a unattainable goal, directly visualise the search for an \textit{(un)stable equilibrium}. These works illustrate the network's unresolvable search for an equilibrium that can never be reached, which result from the network itself being optimising toward a mathematically undefined objective. Revealing the dynamics of a process that is, by construction, interminable. 

One of the goals of this work is to offer a visual and aesthetic means of engaging with the mechanics of gradient descent and the mathematical objectives that undergird contemporary AI systems. The absence of training data acts to foreground the computational processes of training, which is further foregrounded by the training process itself being visualised in the video pieces. The description of the artwork thereby becomes a narrative that guides the viewer toward an understanding of AI through aesthetic experience rather than technical exposition. Viewing the work, in turn, offers an alternative understanding of AI, one grounded in sensation rather than in demonstration. 

Aesthetic experience has long been understood as a catalyst for disruptive and transformation thinking, prompting self-reflection \cite{pelowski2011model} and fostering abstract thinking \cite{mikalonyte2026transformed}. The ambition of this work is to elicit this to some degree. By breaking the normative assumptions of how AI should function, it invites the viewer to question \textit{why} it ends up generating what it does. In doing so, it opens up space for a conceptual understanding of what AI could be, rather than a restatement of what it currently is.

\bibliographystyle{ACM-Reference-Format}
\bibliography{bibliography}

\section*{Appendix}

\end{document}